\pdfoutput=1

\documentclass[12pt,a4paper]{article}

\usepackage{ifthen} 
\newboolean{pdflatex}
\setboolean{pdflatex}{true} 

\newboolean{articletitles}
\setboolean{articletitles}{true} 

\newboolean{uprightparticles}
\setboolean{uprightparticles}{false} 

\newboolean{inbibliography}
\setboolean{inbibliography}{false} 

\newboolean{wordcount}
\setboolean{wordcount}{false} 

\usepackage[top=1in, bottom=1.25in, left=1in, right=1in]{geometry}

\usepackage{microtype}
\usepackage{lineno}  
\usepackage{xspace} 
\usepackage{caption} 

\usepackage{graphicx}  
\usepackage{color}
\usepackage{colortbl}
\graphicspath{{./figs/}} 

\usepackage{amsmath} 
\usepackage{amssymb}
\usepackage{amsfonts}
\usepackage{upgreek} 

\newcommand*\patchAmsMathEnvironmentForLineno[1]{%
\expandafter\let\csname old#1\expandafter\endcsname\csname #1\endcsname
\expandafter\let\csname oldend#1\expandafter\endcsname\csname
end#1\endcsname
 \renewenvironment{#1}%
   {\linenomath\csname old#1\endcsname}%
   {\csname oldend#1\endcsname\endlinenomath}%
}
\newcommand*\patchBothAmsMathEnvironmentsForLineno[1]{%
  \patchAmsMathEnvironmentForLineno{#1}%
  \patchAmsMathEnvironmentForLineno{#1*}%
}
\AtBeginDocument{%
\patchBothAmsMathEnvironmentsForLineno{equation}%
\patchBothAmsMathEnvironmentsForLineno{align}%
\patchBothAmsMathEnvironmentsForLineno{flalign}%
\patchBothAmsMathEnvironmentsForLineno{alignat}%
\patchBothAmsMathEnvironmentsForLineno{gather}%
\patchBothAmsMathEnvironmentsForLineno{multline}%
\patchBothAmsMathEnvironmentsForLineno{eqnarray}%
}

\usepackage{hyperref}    
\usepackage[all]{hypcap} 

\usepackage{xspace} 
\usepackage{upgreek}

\def\lhcb {\mbox{LHCb}\xspace}

\def\MagUp {\mbox{\em Mag\kern -0.05em Up}\xspace}

\ifthenelse{\boolean{uprightparticles}}%
{

 \def\Pmu         {\ensuremath{\upmu}\xspace}

 \def\Ppi         {\ensuremath{\uppi}\xspace}                 
                  
 \def\Prho        {\ensuremath{\uprho}\xspace}

 \def\Pphi        {\ensuremath{\upphi}\xspace}

 \def\Pomega      {\ensuremath{\upomega}\xspace}                 

 \def\PDelta      {\ensuremath{\Delta}\xspace}                 
 \def\PXi      {\ensuremath{\Xi}\xspace}                 
 \def\PLambda      {\ensuremath{\Lambda}\xspace}                 
 \def\PSigma      {\ensuremath{\Sigma}\xspace}                 
 \def\POmega      {\ensuremath{\Omega}\xspace}                 
 \def\PUpsilon      {\ensuremath{\Upsilon}\xspace}                 
 
 \def\PB      {\ensuremath{\mathrm{B}}\xspace}                 
                  
 \def\PD      {\ensuremath{\mathrm{D}}\xspace}

 \def\PK      {\ensuremath{\mathrm{K}}\xspace}

 \def\Pc      {\ensuremath{\mathrm{c}}\xspace}

 \def\Pi      {\ensuremath{\mathrm{i}}\xspace}

 \def\Pp      {\ensuremath{\mathrm{p}}\xspace}

 \def\Pu      {\ensuremath{\mathrm{u}}\xspace}

}
{

 \def\Pmu         {\ensuremath{\mu}\xspace}

 \def\Ppi         {\ensuremath{\pi}\xspace}                 
                  
 \def\Prho        {\ensuremath{\rho}\xspace}

 \def\Pphi        {\ensuremath{\phi}\xspace}

 \def\Pomega      {\ensuremath{\omega}\xspace}                 
 \mathchardef\PDelta="7101
 \mathchardef\PXi="7104
 \mathchardef\PLambda="7103
 \mathchardef\PSigma="7106
 \mathchardef\POmega="710A
 \mathchardef\PUpsilon="7107
                  
 \def\PB      {\ensuremath{B}\xspace}                 
                  
 \def\PD      {\ensuremath{D}\xspace}

 \def\PK      {\ensuremath{K}\xspace}

 \def\Pc      {\ensuremath{c}\xspace}

 \def\Pi      {\ensuremath{i}\xspace}

 \def\Pp      {\ensuremath{p}\xspace}

 \def\Pu      {\ensuremath{u}\xspace}

}

\makeatletter
\ifcase \@ptsize \relax
  \newcommand{\miniscule}{\@setfontsize\miniscule{4}{5}}
\or
  \newcommand{\miniscule}{\@setfontsize\miniscule{5}{6}}
\or
  \newcommand{\miniscule}{\@setfontsize\miniscule{5}{6}}
\fi
\makeatother

\DeclareRobustCommand{\optbar}[1]{\shortstack{{\miniscule (\rule[.5ex]{1.25em}{.18mm})}
  \\ [-.7ex] $#1$}}

\def\mup        {{\ensuremath{\Pmu^+}}\xspace}
\def\mun        {{\ensuremath{\Pmu^-}}\xspace} 

\def\uquark    {{\ensuremath{\Pu}}\xspace}

\def\cquark    {{\ensuremath{\Pc}}\xspace}

\def\pion   {{\ensuremath{\Ppi}}\xspace}

\def\pip    {{\ensuremath{\pion^+}}\xspace}
\def\pim    {{\ensuremath{\pion^-}}\xspace}

\def\rhomeson {{\ensuremath{\Prho}}\xspace}
\def\rhoz     {{\ensuremath{\rhomeson^0}}\xspace}

\def\kaon    {{\ensuremath{\PK}}\xspace}
  \def\Kbar    {{\kern 0.2em\overline{\kern -0.2em \PK}{}}\xspace}

\def\KorKbar    {\kern 0.18em\optbar{\kern -0.18em K}{}\xspace}

\def\Kp      {{\ensuremath{\kaon^+}}\xspace}
\def\Km      {{\ensuremath{\kaon^-}}\xspace}

\newcommand{\phiz}{\ensuremath{\Pphi}\xspace}
\newcommand{\omegaz}{\ensuremath{\Pomega}\xspace}

  \def\Dbar    {{\kern 0.2em\overline{\kern -0.2em \PD}{}}\xspace}
\def\D       {{\ensuremath{\PD}}\xspace}

\def\DorDbar    {\kern 0.18em\optbar{\kern -0.18em D}{}\xspace}
\def\Dz      {{\ensuremath{\D^0}}\xspace}

\def\Dstarp  {{\ensuremath{\D^{*+}}}\xspace}

\def\Bbar    {{\ensuremath{\kern 0.18em\overline{\kern -0.18em \PB}{}}}\xspace}

\def\BorBbar    {\kern 0.18em\optbar{\kern -0.18em B}{}\xspace}

  \def\Y#1S{\ensuremath{\PUpsilon{(#1S)}}\xspace}

\def\proton      {{\ensuremath{\Pp}}\xspace}

\def\Lbar        {{\ensuremath{\kern 0.1em\overline{\kern -0.1em\PLambda}}}\xspace}
\def\LorLbar    {\kern 0.18em\optbar{\kern -0.18em \PLambda}{}\xspace}

\def\BF         {{\ensuremath{\mathcal{B}}}\xspace}

\def\to                 {\ensuremath{\rightarrow}\xspace}

\newcommand{\dm}{{\ensuremath{\Delta m}}\xspace}

\def\AT#1     {\ensuremath{A_{\mathrm{T}}^{#1}}\xspace}           

\def\C#1      {\ensuremath{\mathcal{C}_{#1}}\xspace}                       
\def\Cp#1     {\ensuremath{\mathcal{C}_{#1}^{'}}\xspace}                    
\def\Ceff#1   {\ensuremath{\mathcal{C}_{#1}^{\mathrm{(eff)}}}\xspace}        
\def\Cpeff#1  {\ensuremath{\mathcal{C}_{#1}^{'\mathrm{(eff)}}}\xspace}       
\def\Ope#1    {\ensuremath{\mathcal{O}_{#1}}\xspace}                       
\def\Opep#1   {\ensuremath{\mathcal{O}_{#1}^{'}}\xspace}                    

\newcommand{\tev}{\ifthenelse{\boolean{inbibliography}}{\ensuremath{~T\kern -0.05em eV}}{\ensuremath{\mathrm{\,Te\kern -0.1em V}}}\xspace}
\newcommand{\gev}{\ensuremath{\mathrm{\,Ge\kern -0.1em V}}\xspace}
\newcommand{\mev}{\ensuremath{\mathrm{\,Me\kern -0.1em V}}\xspace}
\newcommand{\kev}{\ensuremath{\mathrm{\,ke\kern -0.1em V}}\xspace}
\newcommand{\ev}{\ensuremath{\mathrm{\,e\kern -0.1em V}}\xspace}
\newcommand{\gevc}{\ensuremath{{\mathrm{\,Ge\kern -0.1em V\!/}c}}\xspace}
\newcommand{\mevc}{\ensuremath{{\mathrm{\,Me\kern -0.1em V\!/}c}}\xspace}
\newcommand{\gevcc}{\ensuremath{{\mathrm{\,Ge\kern -0.1em V\!/}c^2}}\xspace}
\newcommand{\gevgevcccc}{\ensuremath{{\mathrm{\,Ge\kern -0.1em V^2\!/}c^4}}\xspace}
\newcommand{\mevcc}{\ensuremath{{\mathrm{\,Me\kern -0.1em V\!/}c^2}}\xspace}

\def\invfb   {\ensuremath{\mbox{\,fb}^{-1}}\xspace}

\def\gsim{{~\raise.15em\hbox{$>$}\kern-.85em
          \lower.35em\hbox{$\sim$}~}\xspace}
\def\lsim{{~\raise.15em\hbox{$<$}\kern-.85em
          \lower.35em\hbox{$\sim$}~}\xspace}

\def\pt         {\mbox{$p_{\mathrm{ T}}$}\xspace}

\def\tell1  {TELL1\xspace}
\def\ukl1   {UKL1\xspace}

\usepackage{cite} 
\usepackage{mciteplus}

\usepackage{longtable} 

\usepackage{color}

\newcommand{\hhp}{\ensuremath{h^+h^{({\mkern-1mu\prime})-}}\xspace}
\newcommand{\hhmm}{\ensuremath{h^+h^-\mu^+\mu^-}\xspace}
\newcommand{\kpmm}{\ensuremath{\Km\pip\mu^+\mu^-}\xspace}
\newcommand{\kkmm}{\ensuremath{\Kp\Km\mu^+\mu^-}\xspace}
\newcommand{\ppmm}{\ensuremath{\pip\pim\mu^+\mu^-}\xspace}
\newcommand{\Dhhpmm}{\ensuremath{\Dz\to\hhp\mu^+\mu^-}\xspace}
\newcommand{\Dhhmm}{\ensuremath{\Dz\to h^+h^-\mu^+\mu^-}\xspace}
\newcommand{\Dkkmm}{\ensuremath{\Dz\to\Kp\Km\mu^+\mu^-}\xspace}
\newcommand{\Dppmm}{\ensuremath{\Dz\to\pip\pim\mu^+\mu^-}\xspace}
\newcommand{\Dkpmm}{\ensuremath{\Dz\to\Km\pip[\mu^+\mu^-]_{\rhoz/\omegaz}}\xspace}
\newcommand{\Dppmmphi}{\ensuremath{\Dz\to\pip\pim[\mup\mun]_\phi}\xspace}
\newcommand{\Dppmmrho}{\ensuremath{\Dz\to\pip\pim[\mup\mun]_{\rhoz/\omegaz}}\xspace}

\newcommand{\mypaperversion}{}

\newcommand{\mytitle}{Observation of \Dz meson decays to \ppmm and \kkmm final states}
\newcommand{\mylhcbpapernumber}{LHCb-PAPER-2017-019}
\newcommand{\mycernpapernumber}{CERN-EP-2017-167}

\newcommand{\BFppmm}{\ensuremath{9.64}\xspace}
\newcommand{\BFppmmStat}{\ensuremath{0.48}\xspace}
\newcommand{\BFppmmSyst}{\ensuremath{0.51}\xspace}
\newcommand{\BFppmmNorm}{\ensuremath{0.97}\xspace}
\newcommand{\BFppmmUnit}{\ensuremath{\times10^{-7}}\xspace}
\newcommand{\BFkkmm}{\ensuremath{1.54}\xspace}
\newcommand{\BFkkmmStat}{\ensuremath{0.27}\xspace}
\newcommand{\BFkkmmSyst}{\ensuremath{0.09}\xspace}
\newcommand{\BFkkmmNorm}{\ensuremath{0.16}\xspace}
\newcommand{\BFkkmmUnit}{\ensuremath{\times10^{-7}}\xspace}
\newcommand{\BFkpmm}{\ensuremath{4.17}\xspace}

\newcommand{\BFkpmmTot}{\ensuremath{0.42}\xspace}
\newcommand{\BFkpmmUnit}{\ensuremath{\times10^{-6}}\xspace}

\newcommand{\mmumu}{\ensuremath{m(\mup\mun)}\xspace}
\newcommand{\mD}{\ensuremath{m(\Dz)}\xspace}

\begin{document}

\renewcommand{\thefootnote}{\fnsymbol{footnote}}
\setcounter{footnote}{1}


\begin{titlepage}
\pagenumbering{roman}

\vspace*{-1.5cm}
\centerline{\large EUROPEAN ORGANIZATION FOR NUCLEAR RESEARCH (CERN)}
\vspace*{1.5cm}
\noindent
\begin{tabular*}{\linewidth}{lc@{\extracolsep{\fill}}r@{\extracolsep{0pt}}}
\vspace*{-1.7cm}\mbox{\!\!\!\includegraphics[width=.14\textwidth]{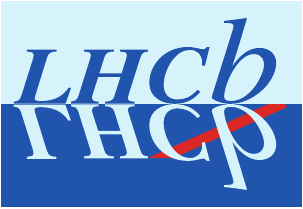}} & & \\
 & & \mycernpapernumber \\  
 & & \mylhcbpapernumber \\  
 & & July 26, 2017 \\
 & & \mypaperversion \\
\end{tabular*}

\vspace*{4.0cm}

{\normalfont\bfseries\boldmath\huge
\begin{center}
\mytitle
\end{center}
}

\vspace*{1.5cm}

\begin{center}
The LHCb collaboration\footnote{Authors are listed at the end of this Letter.}
\end{center}

\vspace{\fill}

\begin{abstract}
\noindent The first observation of the \Dppmm and \Dkkmm decays is reported using a sample of proton-proton collisions collected by LHCb at a center-of-mass energy of 8\tev, and corresponding to 2\invfb of integrated luminosity. The corresponding branching fractions are measured using as normalization the decay \Dkpmm, where the two muons are consistent with coming from the decay of a $\rhoz$ or $\omegaz$ meson. The results are $\BF(\Dppmm)=(\BFppmm\pm\BFppmmStat\pm\BFppmmSyst\pm\BFppmmNorm)\BFppmmUnit$ and $\BF(\Dkkmm)=(\BFkkmm\pm\BFkkmmStat\pm\BFkkmmSyst\pm\BFkkmmNorm)\BFkkmmUnit$, where the uncertainties are statistical, systematic, and due to the limited knowledge of the normalization branching fraction. The dependence of the branching fraction on the dimuon mass is also investigated.
\end{abstract}

\vspace*{2.0cm}

\begin{center}
  Published in Phys.~Rev.~Lett.~119~(2017)~181805
\end{center}

\vspace{\fill}

{\footnotesize 
\centerline{\copyright~CERN on behalf of the \lhcb collaboration, licence \href{http://creativecommons.org/licenses/by/4.0/}{CC-BY-4.0}.}}
\vspace*{2mm}

\end{titlepage}


\newpage
\setcounter{page}{2}
\mbox{~}

\cleardoublepage


\renewcommand{\thefootnote}{\arabic{footnote}}
\setcounter{footnote}{0}


\pagestyle{plain} 
\setcounter{page}{1}
\pagenumbering{arabic}

\begin{figure}[b!]
\centering
\includegraphics[width=0.8\textwidth]{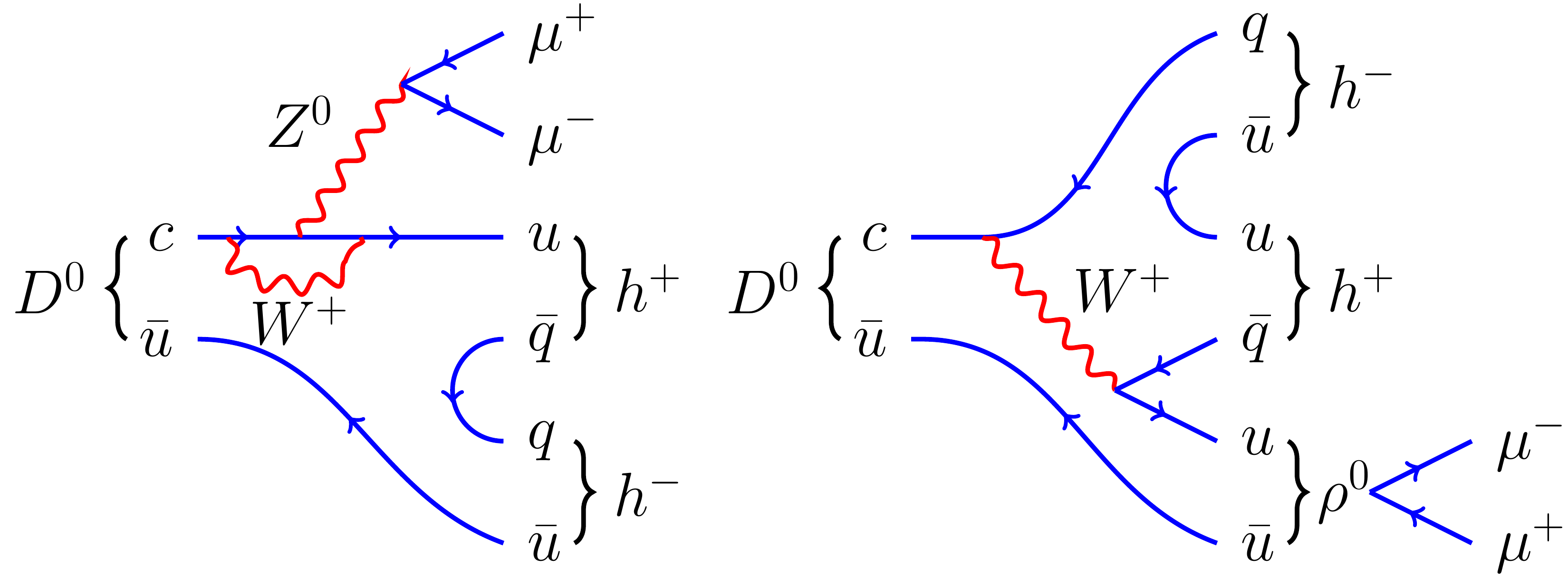}\\
\caption{Example diagrams describing the (left) short- and (right) long-distance contributions to \Dhhmm decays, where $q=d,s$ and $h=\pi,K$.\label{fig:diagrams}}
\end{figure}

Decays of charm hadrons into final states containing dimuon pairs may proceed via the {\em short-distance} $\cquark\to\uquark\mup\mun$ flavor-changing neutral-current process, which in the standard model can only occur through electroweak-loop amplitudes that are highly suppressed by the Glashow-Iliopoulos-Maiani mechanism~\cite{GIM}. If dominated by these short-distance contributions, the inclusive $\D\to X\mup\mun$ branching fraction, where $X$ represents one or more hadrons, is predicted to be $\mathcal{O}(10^{-9})$~\cite{PaulBigi:2011} and can be greatly enhanced by the presence of new particles, making these decays interesting for searches for physics beyond the standard model. However, {\em long-distance} contributions occur through tree-level amplitudes involving intermediate resonances, such as \mbox{$\D\to XV(\to\mup\mun)$}, where $V$ represents a \rhoz, \omegaz or \phiz vector meson, and can increase the standard model branching fraction up to $\mathcal{O}(10^{-6})$~\cite{Fajfer:2007,PaulBigi:2011,Cappiello}. The sensitivity to the short-distance amplitudes is greatest for dimuon masses away from resonances, though resonances populate the entire dimuon-mass spectrum due to their long tails. Additional discrimination between short- and long-distance contributions can be gained by studying angular distributions and charge-parity-conjugation asymmetries, which in scenarios beyond the standard model could be as large as $\mathcal{O}(1\%)$~\cite{Bigi:2012,Paul:2012ab,Fajfer:2015mia,Cappiello,Fajfer:2005ke,deBoer:2015boa}. Decays of \Dz mesons to four-body final states  (Fig.~\ref{fig:diagrams}) are particularly interesting in this respect as they give access to a variety of angular distributions. These decays were searched for by the Fermilab E791 collaboration and upper limits were set on the branching fractions in the range $10^{-5}$--$10^{-4}$ at the 90\% confidence level (CL)~\cite{E791}. More recently, a search for nonresonant \Dppmm decays (the inclusion of charge-conjugate decays is implied) was performed by the LHCb collaboration using 7\tev\ \proton\proton-collision data corresponding to 1\invfb of integrated luminosity~\cite{LHCb-PAPER-2013-050}. An upper limit of $5.5\times10^{-7}$ at the 90\% CL was set on the branching fraction due to short-distance contributions, assuming a phase-space decay.

This Letter reports the first observation of \Dppmm and \Dkkmm decays using data collected by the \lhcb experiment in 2012 at a center-of-mass energy $\sqrt{s}=8\tev$ and corresponding to an integrated luminosity of 2\invfb. The analysis is performed using \Dz mesons originating from $\Dstarp\to\Dz\pip$ decays, with the \Dstarp meson produced directly at the primary \proton\proton-collision vertex (PV). The small phase space available in this decay allows for a large background rejection, which compensates for the reduction in signal yield compared to inclusively produced \Dz mesons. The signal is studied in regions of dimuon mass, \mmumu, defined according to the known resonances. For \Dppmm decays these regions are: (low-mass) $<525\mevcc$, ($\eta$) $525$--$565\mevcc$, ($\rhoz/\omegaz$) $565$--$950\mevcc$, ($\phi$) $950$--$1100\mevcc$, and (high-mass) $>1100\mevcc$. The same regions are considered for \Dkkmm decays, with the exception of the $\phi$ and high-mass regions, which are not present because of the reduced phase space, and the $\rhoz/\omegaz$ region, which extends from 565\mevcc up to the kinematic limit. In the regions where a signal is observed a measurement of the branching fraction is provided, otherwise 90\% and 95\% CL upper limits are set; no attempt is made to distinguish between the short- and long-distance contributions in each dimuon-mass region. The branching fraction is measured using as a normalization the \Dkpmm decay in the dimuon-mass range $675$--$875\mevcc$, where the contribution from the $\rhoz/\omegaz\to\mup\mun$ decay is dominant. The \Dkpmm branching fraction was recently measured to be $(\BFkpmm\pm\BFkpmmTot)\BFkpmmUnit$~\cite{LHCb-PAPER-2015-043} and provides a more precise normalization than that used in the previous LHCb search~\cite{LHCb-PAPER-2013-050}.

The \lhcb detector is a single-arm forward spectrometer~\cite{Alves:2008zz,LHCb-DP-2014-002}. It includes a high-precision tracking system consisting of a silicon-strip vertex detector surrounding the \proton\proton-interaction region, a large-area silicon-strip detector located upstream of a dipole magnet with a bending power of about 4\,Tm, and three stations of silicon-strip detectors and straw drift tubes placed downstream of the magnet. Particle identification is provided by two ring-imaging Cherenkov detectors, an electromagnetic and a hadronic calorimeter, and a muon system composed of alternating layers of iron and multiwire proportional chambers. 

Events are selected online by a trigger that consists of a hardware stage, which is based on information from the calorimeter and muon systems, followed by a software stage, which applies a full event reconstruction~\cite{LHCb-DP-2012-004}. The hardware trigger requires the presence in the event of a muon with transverse momentum, \pt, exceeding 1.76\gevc. A first stage of the software trigger selects events with a charged particle of $\pt>1.6\gevc$ and significant impact parameter, defined as the minimum distance of the particle trajectory from any PV, or alternatively with $\pt>1\gevc$ if the particle has associated hits in the muon system. In a second stage of the software trigger, dedicated algorithms select candidate \Dhhpmm decays, where $h$ is either a kaon or a pion, from combinations of four tracks, each having momentum $p>3\gevc$ and $\pt>0.5\gevc$, that form a secondary vertex separated from any PV. Two oppositely charged particles are required to leave hits in the muon system and the scalar sum of their \pt is required to exceed 3\gevc. The mass of the \Dz candidate, \mD, has to be in the range $1800$--$1940\mevcc$ and its momentum must be aligned with the vector connecting the primary and secondary vertices.

In the offline analysis, \Dz candidates satisfying the trigger requirements are further selected through particle-identification criteria placed on their decay products. They are then combined with a charged particle originating from the same PV and having $\pt>120\mevc$, to form a $\Dstarp\to\Dz\pip$ candidate. When more than one PV is reconstructed, the one with respect to which the \Dz candidate has the lowest impact-parameter significance is chosen. The vertex formed by the \Dz and \pip mesons is constrained to coincide with the PV and the difference between the \Dstarp and \Dz masses, \dm, is required to be in the range $144.5$--$146.5$\mevcc. A multivariate selection based on a boosted decision tree (BDT)~\cite{Breiman,Roe} with gradient boosting~\cite{TMVA} is then used to suppress background from combinations of unrelated charged particles. The features used by the BDT to discriminate signal from this {\em combinatorial} background are as follows: the momentum and transverse momentum of the pion from the \Dstarp decay, the smallest impact parameter of the \Dz decay products with respect to the PV, the angle between the \Dz momentum and the vector connecting the primary and secondary vertices, the quality of the secondary vertex, its separation from the PV, and its separation from any other track not forming the \Dstarp candidate. The BDT is trained separately for \mbox{\Dppmm} and \Dkkmm decays, due to their different kinematic properties, using simulated~\cite{LHCb-PROC-2011-006, LHCb-PROC-2010-056} decays as signal and data candidates with \mD between 1890 and 1940\mevcc as background. To minimize biases on the background classification, the training samples are further randomly split into two disjoint subsamples. The classifier trained on one sample is applied to the other, and vice versa. Another source of background is due to the hadronic four-body decays $\Dz\to\pip\pim\pip\pim$ and $\Dz\to\Kp\Km\pip\pim$, where two pions are misidentified as muons. The misidentification occurs mainly when the pions decay in flight into a muon and an undetected neutrino. Although this process is relatively rare, the large branching fractions of the hadronic modes produce a peaking background which is partially suppressed by a multivariate muon-identification discriminant that combines the information from the Cherenkov detectors, the calorimeters and the muon chambers. Thresholds on the BDT response and on the muon-identification discriminant are optimized simultaneously by maximizing ${\epsilon_\text{\hhmm}}/({5/2 + \sqrt{N_\text{bkg}}})$~\cite{Punzi:2003bu}, where $\epsilon_\text{\hhmm}$ is the signal efficiency and $N_\text{bkg}$ is the sum of the expected combinatorial and peaking background yields in the \mD range $1830$--$1900$\mevcc ({\em signal region}). Candidate \Dkpmm decays are selected using the response of the BDT trained on the \Dppmm signal, when they are used as normalization for the measurement of $\BF(\Dppmm)$, and that of the BDT trained on the \Dkkmm signal, when used as normalization for $\BF(\Dkkmm)$. After selection, a few percent of the events contain multiple candidates, of which only one is randomly selected if they share at least one final-state particle. To avoid potential biases on the measured quantities, candidate decays in the \mD signal region are examined only after the analysis procedure has been finalized, with the exception of those populating the $\rhoz/\omega$ and $\phi$ dimuon-mass regions of the \Dppmm sample.

\begin{figure}[tb]
\centering
\includegraphics[width=0.55\textwidth]{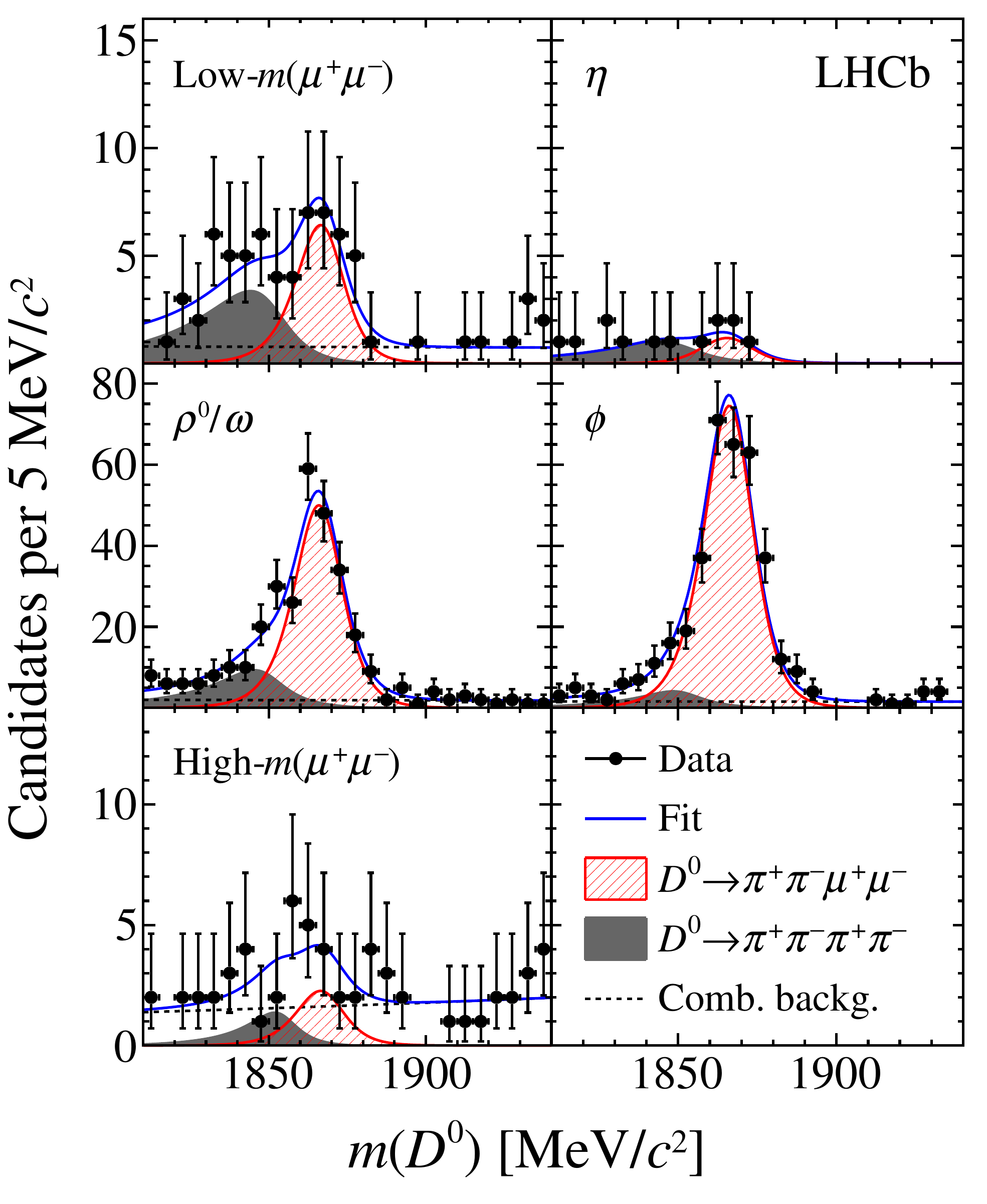}\\
\caption{Distributions of \mD for the \Dppmm candidates in the low-\mmumu, $\eta$, $\rhoz/\omegaz$, $\phi$ and high-\mmumu regions. Fit projections are overlaid.\label{fig:massfits-ppmm}}
\end{figure}

\begin{figure}[tb]
\centering
\includegraphics[width=0.55\textwidth]{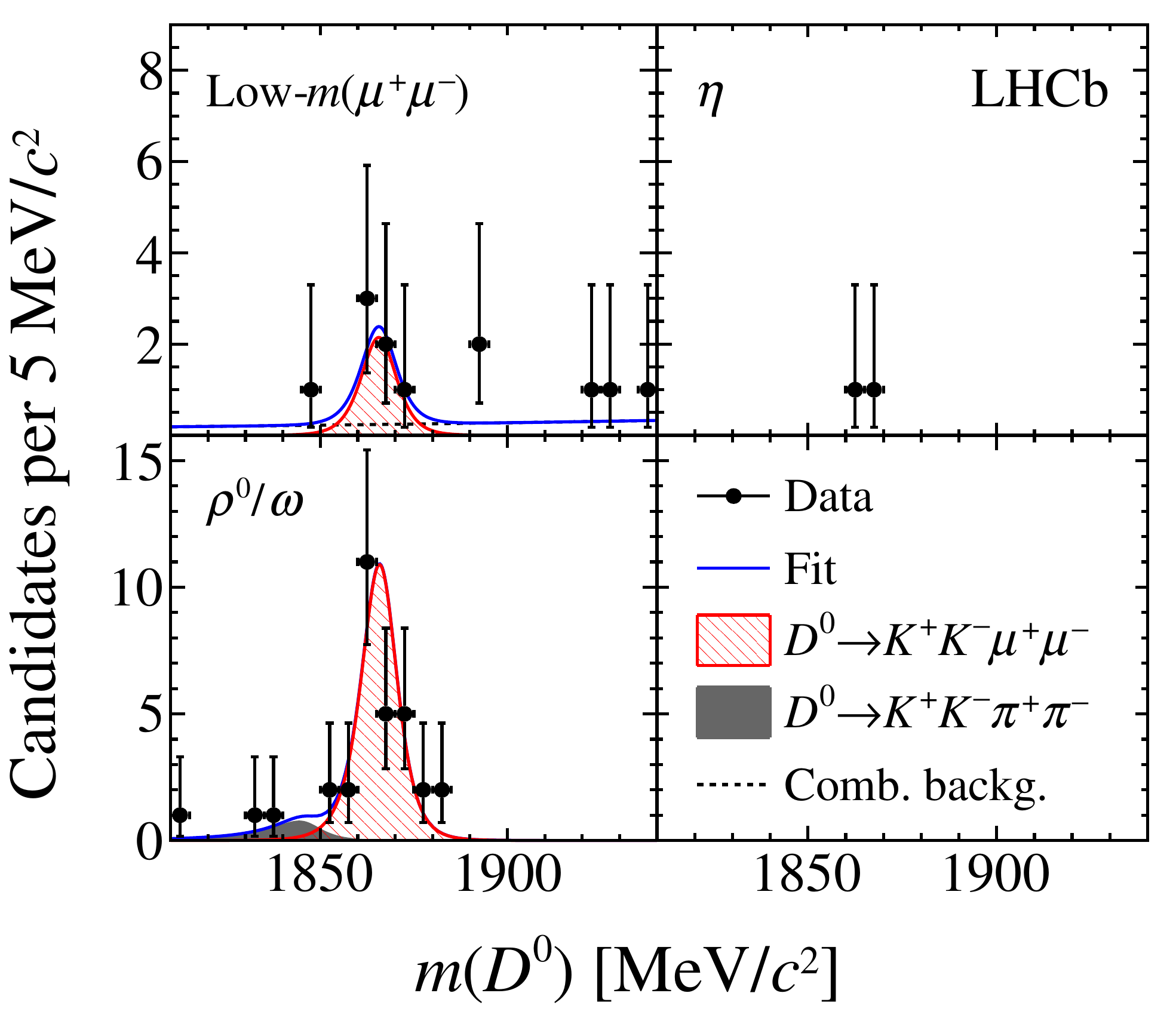}\\
\caption{Distributions of \mD for the \Dkkmm candidates in the low-\mmumu,  $\eta$ and $\rhoz/\omegaz$ regions. Fit projections are overlaid. No fit is performed in the $\eta$ region, where only two candidates are observed.\label{fig:massfits-kkmm}}
\end{figure}

The \Dppmm and \Dkkmm signal yields are measured with unbinned extended maximum likelihood fits to the \mD distributions (Figs.~\ref{fig:massfits-ppmm} and \ref{fig:massfits-kkmm}, respectively). The fits include three components: signal, peaking background from misidentified hadronic decays, and combinatorial background. The signal is described with a Johnson's $S_U$ distribution~\cite{johnson} with parameters determined from simulation. To account for known differences between data and simulation, the means and widths of the signal distributions are corrected using scaling factors adjusted on the normalization channel. The mass shape of the peaking background is determined using separate data samples of $\Dz\to\hhp\pip\pim$ decays where the \Dz mass is calculated assigning the muon-mass hypothesis to two oppositely charged pions. The combinatorial background is described by an exponential function, which is determined from data candidates with \dm between 150 and 160\mevcc that fail the BDT selection. All shape parameters are fixed and only the yields are allowed to vary in the fits, which are performed separately in each \mmumu range.

\begin{table}[tb]
\centering
\caption{Yields of (top) \Dppmm and (bottom) \Dkkmm signal decays, their significance with respect to the background-only hypothesis, and ratio of efficiencies between signal and normalization decays ($R_\epsilon^i$) for each dimuon-mass region. The yield and the significance ($\mathcal{S}$) are not reported for the $\eta$ region of \Dkkmm, where only two candidates are observed.}\label{tab:yields-effs}
\ifthenelse{\boolean{wordcount}}{}{%
\begin{tabular}{lr@{--}lr@{\,$\pm$\,}lrr@{\,$\pm$\,}l}
\hline
\multicolumn{8}{c}{\Dppmm}\\
\mmumu region & \multicolumn{2}{c}{[$\mevcc$]} & \multicolumn{2}{c}{Yield} & \multicolumn{1}{c}{$\mathcal{S}$} & \multicolumn{2}{c}{$R_\epsilon^i$} \\
Low mass        & \multicolumn{2}{c}{$<525$}  &  27 &  6 & $5.4\sigma$ & 0.73 & 0.04 \\
$\eta$          & 525 & 565                   &   5 &  3 & $2.5\sigma$ & 0.84 & 0.07 \\
$\rhoz/\omegaz$ & 565 & 950                   & 208 & 17 &  $18\sigma$ & 1.08 & 0.05 \\
$\phi$          & 950 & 1100                  & 312 & 20 &  $23\sigma$ & 1.45 & 0.07 \\
High mass       & \multicolumn{2}{c}{$>1100$} &   9 &  6 & $1.6\sigma$ &  1.5 & 0.1 \\
\hline
\multicolumn{8}{c}{\Dkkmm}\\
\mmumu region & \multicolumn{2}{c}{[$\mevcc$]} & \multicolumn{2}{c}{Yield} & \multicolumn{1}{c}{$\mathcal{S}$} & \multicolumn{2}{c}{$R_\epsilon^i$} \\
Low mass        & \multicolumn{2}{c}{$<525$}  &  5 &  3 & $3.1\sigma$ & 0.49 & 0.03 \\
$\eta$          & 525 & 565                   & \multicolumn{2}{c}{$\,\,\,$--} & -- & 0.53 & 0.04 \\
$\rhoz/\omegaz$ & \multicolumn{2}{c}{$>565$}  & 29 &  5 & $8.1\sigma$ & 0.55 & 0.03 \\
\hline
\end{tabular}}
\end{table}

The resulting signal yields are reported in Table~\ref{tab:yields-effs}. No fit is performed in the $\eta$ region of the \Dkkmm dimuon-mass spectrum, where only two candidates are observed. An excess of candidates with respect to the background-only hypothesis is seen with a significance above three standard deviations in all dimuon-mass ranges with the exception of the $\eta$ region of both decays and the high-\mmumu region of \Dppmm. The significances are determined from the change in likelihood from fits with and without the signal component. 

The signal yields, $N^i_{\hhmm}$, in each \mmumu range $i$ are converted into branching fractions using
\ifthenelse{\boolean{wordcount}}{}{%
\begin{equation}
\BF^i(\Dhhmm) = \frac{N_{\hhmm}^i\, \BF(\Dkpmm)}{R^i_\epsilon\,N_{\kpmm}},
\end{equation}
}
where $N_{\kpmm}$ is the yield of the normalization mode, which is determined to be \mbox{$1971\pm51$} ($1806\pm48$) after the selection optimized for \mbox{\Dppmm} (\mbox{\Dkkmm}) decays. The ratios of geometrical acceptances, and reconstruction and selection efficiencies of the signal relative to the normalization decays, \mbox{$R_\epsilon^i=\epsilon_{\hhmm}^i/\epsilon_{\kpmm}$}, are reported in Table~\ref{tab:yields-effs}. They are determined using simulated events and corrected to account for known differences between data and simulation. In particular, particle-identification and hardware-trigger efficiencies are measured from control channels in data.

Systematic uncertainties affect the determination of the signal and normalization yields, and of the efficiency ratio. For the determination of the yields, effects due to uncertainties on the \mD shapes are investigated. A possible dependence on the decay mode or on the \mmumu range of the scaling factors, used to account for data-simulation differences, is quantified using fits to the \Dppmmphi and \Dppmmrho data and is found to be negligible. To assess the impact of $\pi\to\mu\nu$ decays in flight, alternative shapes are tested for the $\Dz\to\hhp\pip\pim$ background by changing the muon-identification and the \pt requirements on the misidentified pions. The largest observed variation in the ratio of \Dppmmphi to \Dkpmm yields (1.4\%) is assigned as a systematic uncertainty for both \hhmm modes and all dimuon-mass ranges. Changes in the shape of the peaking background introduced by the different trigger requirements used to select the hadronic decays are negligible. The fit to the data is repeated using alternative descriptions of the combinatorial background, determined from data sidebands defined by different BDT and \dm requirements, and results in negligible variations of the signal and normalization yields.

Systematic uncertainties affecting the efficiency ratio include data-simulation differences that are not accounted for and limitations in the data-driven methods used to determine the particle-identification and trigger efficiencies. The signal decays are simulated with an incoherent sum of resonant and nonresonant dimuon and dihadron components, while the resonant structure in data is unknown. A systematic uncertainty of 3.4\% on the signal efficiency is determined by varying the relative fractions of these components. A systematic uncertainty of 1.0\% on the efficiency ratio is assigned due to the criteria used in simulation to match the reconstructed and generated particles. Muon- and hadron-identification efficiencies are determined from data by weighting the kinematic properties of the calibration samples to match those of the signal samples. Variations of the choice of the binning scheme used in the weighting procedure change the efficiency ratio by up to 0.8\%, which is taken as systematic uncertainty. The data-driven method that evaluates the hardware-trigger efficiency ratio is validated in simulation to be unbiased within 1.3\%, which is assigned as a systematic uncertainty. The efficiencies of the BDT requirement for the simulated normalization and \Dppmmphi decays are compared to those obtained from background-subtracted data. A difference in the efficiency ratio of 1.3\% is observed and assigned as systematic uncertainty.

Finally, the statistical uncertainty on the normalization yield introduces a relative uncertainty of $2.6\%$ ($2.7\%$), which is propagated to the systematic uncertainty on the \Dppmm (\Dkkmm) branching fractions.

\begin{table}[tb]
\centering
\caption{Branching fractions of (top) \Dppmm and (bottom) \Dkkmm decays in different ranges of dimuon mass, where the uncertainties are statistical, systematic and due to the limited knowledge of the normalization branching fraction. The reported upper limits correspond to 90\% (95\%) CL. The correlations between the various dimuon-mass ranges are reported in the supplemental material \cite{supplemental}.\label{tab:BF}}
\ifthenelse{\boolean{wordcount}}{}{%
\begin{tabular}{lr@{--}lc}
\hline
\multicolumn{4}{c}{\Dppmm}\\
\mmumu region & \multicolumn{2}{c}{[$\mevcc$]} & $\BF$ [$10^{-8}$]\\
Low mass        & \multicolumn{2}{c}{$<525$} & $\enspace7.8\pm 1.9 \pm 0.5 \pm 0.8$\\
$\eta$          & 525 & 565  & $<2.4\,(2.8)$\\
$\rhoz/\omegaz$ & 565 & 950  & $40.6\pm 3.3 \pm 2.1 \pm 4.1$\\
$\phi$          & 950 & 1100 & $45.4\pm 2.9 \pm 2.5 \pm 4.5$\\
High mass       & \multicolumn{2}{c}{$>1100$} & $<2.8\,(3.3)$ \\
\hline
\multicolumn{4}{c}{\Dkkmm}\\
\mmumu region & \multicolumn{2}{c}{[$\mevcc$]} & $\BF$ [$10^{-8}$]\\
Low mass        &\multicolumn{2}{c}{$<525$} & $\enspace2.6 \pm  1.2 \pm 0.2 \pm 0.3$\\
$\eta$          &525 & 565  & $<0.7\,(0.8)$ \\
$\rhoz/\omegaz$ &\multicolumn{2}{c}{$>565$} & $12.0 \pm 2.3  \pm 0.7  \pm 1.2$ \\
\hline
\end{tabular}}
\end{table}

Table~\ref{tab:BF} reports the measured values and upper limits on the \Dppmm and \Dkkmm branching fractions in the various ranges of \mmumu, where the first uncertainty accounts for the statistical component, the second for the systematic, and the third corresponds to the 10\% relative uncertainty on $\BF(\Dkpmm)$~\cite{LHCb-PAPER-2015-043}. The upper limits are derived using a frequentist approach based on a likelihood-ratio ordering method that includes the effects due to the systematic uncertainties~\cite{CLs,Junk:1999kv}. For the $\eta$ region of \Dkkmm, where no fit is performed, the limit is calculated assuming two signal candidates and zero background. Integrating over dimuon mass, and accounting for correlations~\cite{supplemental}, the total branching fractions are measured to be
\begin{align}
\BF(\Dppmm)&=(\BFppmm\pm\BFppmmStat\pm\BFppmmSyst\pm\BFppmmNorm)\BFppmmUnit,\nonumber\\
\BF(\Dkkmm)&=(\BFkkmm\pm\BFkkmmStat\pm\BFkkmmSyst\pm\BFkkmmNorm)\BFkkmmUnit.
\end{align}
The two results have a correlation of $0.497$ and are consistent with the standard model expectations~\cite{Cappiello}.

In summary, a study of the \Dppmm and \Dkkmm decays is performed in ranges of the dimuon mass using \proton\proton collisions collected by the \lhcb experiment at $\sqrt{s}=8\tev$. Significant signal yields are observed for the first time in several dimuon-mass ranges for both decays; the corresponding branching fractions are measured and found to be consistent with the standard model expectations~\cite{Cappiello}. For the dimuon-mass regions where no significant signal is observed, upper limits at 90\% and 95\% CL are set on the branching fraction. The total branching fractions are measured to be $\BF(\Dppmm)=(\BFppmm\pm\BFppmmStat\pm\BFppmmSyst\pm\BFppmmNorm)\BFppmmUnit$ and $\BF(\Dkkmm)=(\BFkkmm\pm\BFkkmmStat\pm\BFkkmmSyst\pm\BFkkmmNorm)\BFkkmmUnit$, where the uncertainties are statistical, systematic, and due to the limited knowledge of the normalization branching fraction. These are the rarest charm-hadron decays ever observed and are expected to provide better sensitivity to short-distance flavor-changing neutral-current contributions to these decays.

\section*{Acknowledgements}

%
%
\noindent We express our gratitude to our colleagues in the CERN
accelerator departments for the excellent performance of the LHC. We
thank the technical and administrative staff at the LHCb
institutes. We acknowledge support from CERN and from the national
agencies: CAPES, CNPq, FAPERJ and FINEP (Brazil); MOST and NSFC (China);
CNRS/IN2P3 (France); BMBF, DFG and MPG (Germany); INFN (Italy); 
NWO (The Netherlands); MNiSW and NCN (Poland); MEN/IFA (Romania); 
MinES and FASO (Russia); MinECo (Spain); SNSF and SER (Switzerland); 
NASU (Ukraine); STFC (United Kingdom); NSF (USA).
We acknowledge the computing resources that are provided by CERN, IN2P3 (France), KIT and DESY (Germany), INFN (Italy), SURF (The Netherlands), PIC (Spain), GridPP (United Kingdom), RRCKI and Yandex LLC (Russia), CSCS (Switzerland), IFIN-HH (Romania), CBPF (Brazil), PL-GRID (Poland) and OSC (USA). We are indebted to the communities behind the multiple open 
source software packages on which we depend.
Individual groups or members have received support from AvH Foundation (Germany),
EPLANET, Marie Sk\l{}odowska-Curie Actions and ERC (European Union), 
Conseil G\'{e}n\'{e}ral de Haute-Savoie, Labex ENIGMASS and OCEVU, 
R\'{e}gion Auvergne (France), RFBR and Yandex LLC (Russia), GVA, XuntaGal and GENCAT (Spain), Herchel Smith Fund, The Royal Society, Royal Commission for the Exhibition of 1851 and the Leverhulme Trust (United Kingdom).

\addcontentsline{toc}{section}{References}
\setboolean{inbibliography}{true}
\bibliographystyle{LHCb}
\bibliography{main,LHCb-PAPER,LHCb-CONF,LHCb-DP,LHCb-TDR,mybib}

\newpage
\clearpage
\section*{Supplemental material}\label{sec:supp-prl}
The correlations between $\BF(\Dppmm)$ and $\BF(\Dkkmm)$ in the dimuon-mass regions are reported in \tablename~\ref{tab:corr-ppmm} and \tablename~\ref{tab:corr-kkmm}, respectively. The correlation between the total branching fractions is $0.497$.

\begin{table}[h]
\centering
\caption{Correlation coefficients between the \Dppmm branching fractions in the dimuon-mass ranges.\label{tab:corr-ppmm}}
\begin{tabular}[c]{lrrrrr}
\hline
\multicolumn{6}{c}{$\BF(\Dppmm)$}\\
\hline
[\mevcc]  & $<525$ & 525--565 & 565--950 & 950--1100 & $>1100$ \\
$<525$    &  1.000 &    0.088 &    0.343 &     0.366 &   0.078 \\
525--565  &        &    1.000 &    0.170 &     0.181 &   0.039 \\
565--950  &        &          &    1.000 &     0.706 &   0.151 \\
950--1100 &        &          &          &     1.000 &   0.161 \\
$>1100$   &        &          &          &           &   1.000 \\
\hline
\end{tabular}
\end{table}

\begin{table}[h]
\centering
\caption{Correlation coefficients between the \Dkkmm branching fractions in the dimuon-mass ranges.\label{tab:corr-kkmm}}
\begin{tabular}[c]{lrrr}
\hline
\multicolumn{4}{c}{$\BF(\Dkkmm)$} \\
\hline
[\mevcc]  & $<525$ & 525--565 & $>565$ \\
$<525$    &  1.000 &    0.027 &  0.123 \\
525--565  &        &    1.000 &  0.059 \\
$>565$    &        &          &  1.000 \\
\hline									
\end{tabular}
\end{table}


\newpage
\centerline{\large\bf LHCb collaboration}
\begin{flushleft}
\small
R.~Aaij$^{40}$,
B.~Adeva$^{39}$,
M.~Adinolfi$^{48}$,
Z.~Ajaltouni$^{5}$,
S.~Akar$^{59}$,
J.~Albrecht$^{10}$,
F.~Alessio$^{40}$,
M.~Alexander$^{53}$,
A.~Alfonso~Albero$^{38}$,
S.~Ali$^{43}$
% ...

{\footnotesize \it
% ...
$ ^{5}$Clermont Universit{\'e}, Universit{\'e} Blaise Pascal, CNRS/IN2P3, LPC, Clermont-Ferrand, France\\
% ...
$ ^{10}$Fakult{\"a}t Physik, Technische Universit{\"a}t Dortmund, Dortmund, Germany\\
% ...
$ ^{38}$ICCUB, Universitat de Barcelona, Barcelona, Spain\\
$ ^{39}$Universidad de Santiago de Compostela, Santiago de Compostela, Spain\\
$ ^{40}$European Organization for Nuclear Research (CERN), Geneva, Switzerland\\
% ...
$ ^{43}$Nikhef National Institute for Subatomic Physics, Amsterdam, The Netherlands\\
% ...
$ ^{48}$H.H. Wills Physics Laboratory, University of Bristol, Bristol, United Kingdom\\
% ...
$ ^{53}$School of Physics and Astronomy, University of Glasgow, Glasgow, United Kingdom\\
% ...
$ ^{59}$University of Cincinnati, Cincinnati, OH, United States\\
% ...
}
\end{flushleft}

\end{document}